\UseRawInputEncoding 
\documentclass[10pt,aps,prd,twocolumn,superscriptaddress,nofootinbib,floatfix]{revtex4-2}
\usepackage{ragged2e}
\usepackage{float}
\usepackage{amsmath, amssymb, amsfonts}
\usepackage{graphicx}
\usepackage[export]{adjustbox} 
\usepackage{caption}  
\usepackage{subcaption}  
\usepackage{multirow}  
\usepackage{booktabs}  
\usepackage{rotating}  
\usepackage{comment}  
\usepackage{epsfig}  
\usepackage{xcolor}  
 \usepackage[normalem]{ulem}
\usepackage{hyperref}  
\usepackage{cleveref}  
\usepackage{placeins}
\FloatBarrier
\usepackage[left]{lineno}
\usepackage{orcidlink}
\usepackage{comment}

\newcommand{\Albanysuny}{University of Albany, SUNY, Albany, NY 12222, USA}
\newcommand{\Almaty}{Institute of Nuclear Physics at Almaty, Almaty 050032, Kazakhstan
}
\newcommand{\Amsterdam}{University of Amsterdam, NL-1098 XG Amsterdam, The Netherlands}
\newcommand{\Antalya}{Antalya Bilim University, 07190 D\"o{\c s}emealtı/Antalya, Turkey}
\newcommand{\Antananarivo}{University of Antananarivo, Antananarivo 101, Madagascar}
\newcommand{\Antioquia}{University of Antioquia, Medell\'in, Colombia}
\newcommand{\AntonioNarino}{Universidad Antonio Nari\~no, Bogot\'a, Colombia}
\newcommand{\Argonne}{Argonne National Laboratory, Argonne, IL 60439, USA}
\newcommand{\Arizona}{University of Arizona, Tucson, AZ 85721, USA}
\newcommand{\Asuncion}{Universidad Nacional de Asunci\'on, San Lorenzo, Paraguay}
\newcommand{\Athens}{University of Athens, Zografou GR 157 84, Greece}
\newcommand{\Atlantico}{Universidad del Atl\'antico, Barranquilla, Atl\'antico, Colombia}
\newcommand{\Augustana}{Augustana University, Sioux Falls, SD 57197, USA}
\newcommand{\Bern}{University of Bern, CH-3012 Bern, Switzerland}
\newcommand{\Beykent}{Beykent University, Istanbul, Turkey}
\newcommand{\Birmingham}{University of Birmingham, Birmingham B15 2TT, United Kingdom}
\newcommand{\BolognaUniversity}{Universit\`a di Bologna, 40127 Bologna, Italy}
\newcommand{\Boston}{Boston University, Boston, MA 02215, USA}
\newcommand{\Bristol}{University of Bristol, Bristol BS8 1TL, United Kingdom}
\newcommand{\Brookhaven}{Brookhaven National Laboratory, Upton, NY 11973, USA}
\newcommand{\Bucharest}{University of Bucharest, Bucharest, Romania}
\newcommand{\CalBerkeley}{University of California Berkeley, Berkeley, CA 94720, USA}
\newcommand{\CalDavis}{University of California Davis, Davis, CA 95616, USA}
\newcommand{\CalIrvine}{University of California Irvine, Irvine, CA 92697, USA}
\newcommand{\CalLosangeles}{University of California Los Angeles, Los Angeles, CA 90095, USA}
\newcommand{\CalRiverside}{University of California Riverside, Riverside CA 92521, USA}
\newcommand{\CalSantabarbara}{University of California Santa Barbara, Santa Barbara, CA 93106, USA}
\newcommand{\Caltech}{California Institute of Technology, Pasadena, CA 91125, USA}
\newcommand{\Cambridge}{University of Cambridge, Cambridge CB3 0HE, United Kingdom}
\newcommand{\Campinas}{Universidade Estadual de Campinas, Campinas - SP, 13083-970, Brazil}
\newcommand{\CataniaUniversitadi}{Universit\`a di Catania, 2 - 95131 Catania, Italy}
\newcommand{\Catolica}{Universidad Cat\'olica del Norte, Antofagasta, Chile}
\newcommand{\CBPF}{Centro Brasileiro de Pesquisas F\'isicas, Rio de Janeiro, RJ 22290-180, Brazil}
\newcommand{\CEASaclay}{IRFU, CEA, Universit\'e Paris-Saclay, F-91191 Gif-sur-Yvette, France}
\newcommand{\CERN}{CERN, The European Organization for Nuclear Research, 1211 Meyrin, Switzerland}
\newcommand{\Charles}{Institute of Particle and Nuclear Physics of the Faculty of Mathematics and Physics of the Charles University, 180 00 Prague 8, Czech Republic }
\newcommand{\Chicago}{University of Chicago, Chicago, IL 60637, USA}
\newcommand{\ChungAng}{Chung-Ang University, Seoul 06974, South Korea}
\newcommand{\CIEMAT}{CIEMAT, Centro de Investigaciones Energ\'eticas, Medioambientales y Tecnol\'ogicas, E-28040 Madrid, Spain}
\newcommand{\Cincinnati}{University of Cincinnati, Cincinnati, OH 45221, USA}
\newcommand{\Cinvestav}{Centro de Investigaci\'on y de Estudios Avanzados del Instituto Polit\'ecnico Nacional (Cinvestav), Mexico City, Mexico}
\newcommand{\Colima}{Universidad de Colima, Colima, Mexico}
\newcommand{\ColoradoBoulder}{University of Colorado Boulder, Boulder, CO 80309, USA}
\newcommand{\ColoradoState}{Colorado State University, Fort Collins, CO 80523, USA}
\newcommand{\Columbia}{Columbia University, New York, NY 10027, USA}
\newcommand{\conida}{Comisi\'on Nacional de Investigaci\'on y Desarrollo Aeroespacial, Lima, Peru}
\newcommand{\Cti}{Centro de Tecnologia da Informacao Renato Archer, Amarais - Campinas, SP - CEP 13069-901}
\newcommand{\CUSB}{Central University of South Bihar, Gaya, 824236, India
}
\newcommand{\CzechAcademyofSciences}{Institute of Physics, Czech Academy of Sciences, 182 00 Prague 8, Czech Republic}
\newcommand{\CzechTechnical}{Czech Technical University, 115 19 Prague 1, Czech Republic}
\newcommand{\DannecyleVieux}{Laboratoire d'Annecy de Physique des Particules, Universit\'e Savoie Mont Blanc, CNRS, LAPP-IN2P3, 74000 Annecy, France}
\newcommand{\Daresbury}{Daresbury Laboratory, Cheshire WA4 4AD, United Kingdom}
\newcommand{\Dordt}{Dordt University, Sioux Center, IA 51250, USA}
\newcommand{\Drexel}{Drexel University, Philadelphia, PA 19104, USA}
\newcommand{\Duke}{Duke University, Durham, NC 27708, USA}
\newcommand{\Edinburgh}{University of Edinburgh, Edinburgh EH8 9YL, United Kingdom}
\newcommand{\EIA}{Universidad EIA, Envigado, Antioquia, Colombia}
\newcommand{\Eotvos}{E\"otv\"os Lor\'and University, 1053 Budapest, Hungary}
\newcommand{\erciyes}{Erciyes University, Kayseri, Turkey}
\newcommand{\FCULport}{Faculdade de Ci\^encias da Universidade de Lisboa - FCUL, 1749-016 Lisboa, Portugal}
\newcommand{\FederaldeAlfenas}{Universidade Federal de Alfenas, Po{\c c}os de Caldas - MG, 37715-400, Brazil}
\newcommand{\FederaldeGoias}{Universidade Federal de Goias, Goiania, GO 74690-900, Brazil}
\newcommand{\FederaldoABC}{Universidade Federal do ABC, Santo Andr\'e - SP, 09210-580, Brazil}
\newcommand{\FederaldoRio}{Universidade Federal do Rio de Janeiro, Rio de Janeiro - RJ, 21941-901, Brazil}
\newcommand{\Fermi}{Fermi National Accelerator Laboratory, Batavia, IL 60510, USA}
\newcommand{\Ferrarauniv}{University of Ferrara, Ferrara, Italy}
\newcommand{\Florida}{University of Florida, Gainesville, FL 32611-8440, USA}
\newcommand{\Floridastate}{Florida State University, Tallahassee, FL, 32306 USA}
\newcommand{\Fluminense}{Fluminense Federal University, 9 Icara\'i Niter\'oi - RJ, 24220-900, Brazil }
\newcommand{\Genova}{Universit\`a degli Studi di Genova, Genova, Italy}
\newcommand{\Georgian}{Georgian Technical University, Tbilisi, Georgia}
\newcommand{\Granada}{University of Granada \& CAFPE, 18002 Granada, Spain}
\newcommand{\GranSasso}{Gran Sasso Science Institute, L'Aquila, Italy}
\newcommand{\GranSassoLab}{Laboratori Nazionali del Gran Sasso, L'Aquila AQ, Italy}
\newcommand{\Grenoble}{University Grenoble Alpes, CNRS, Grenoble INP, LPSC-IN2P3, 38000 Grenoble, France}
\newcommand{\Guanajuato}{Universidad de Guanajuato, Guanajuato, C.P. 37000, Mexico}
\newcommand{\Harish}{Harish-Chandra Research Institute, Jhunsi, Allahabad 211 019, India}
\newcommand{\Hawaii}{University of Hawaii, Honolulu, HI 96822, USA}
\newcommand{\hkust}{Hong Kong University of Science and Technology, Kowloon, Hong Kong, China}
\newcommand{\Houston}{University of Houston, Houston, TX 77204, USA}
\newcommand{\Hyderabad}{University of  Hyderabad, Gachibowli, Hyderabad - 500 046, India}
\newcommand{\Idaho}{Idaho State University, Pocatello, ID 83209, USA}
\newcommand{\IFIC}{Instituto de F\'isica Corpuscular, CSIC and Universitat de Val\`encia, 46980 Paterna, Valencia, Spain}
\newcommand{\IGFAE}{Instituto Galego de F\'isica de Altas Enerx\'ias, University of Santiago de Compostela, Santiago de Compostela, 15782, Spain}
\newcommand{\ihep}{Institute of High Energy Physics, Chinese Academy of Sciences, Beijing, China}
\newcommand{\Iitk}{Indian Institute of Technology Kanpur, Uttar Pradesh 208016, India}
\newcommand{\Illinoisinstitute}{Illinois Institute of Technology, Chicago, IL 60616, USA}
\newcommand{\Imperial}{Imperial College of Science, Technology and Medicine, London SW7 2BZ, United Kingdom}
\newcommand{\IndGuwahati}{Indian Institute of Technology Guwahati, Guwahati, 781 039, India}
\newcommand{\IndHyderabad}{Indian Institute of Technology Hyderabad, Hyderabad, 502285, India}
\newcommand{\Indiana}{Indiana University, Bloomington, IN 47405, USA}
\newcommand{\INFNBologna}{Istituto Nazionale di Fisica Nucleare Sezione di Bologna, 40127 Bologna BO, Italy}
\newcommand{\INFNCatania}{Istituto Nazionale di Fisica Nucleare Sezione di Catania, I-95123 Catania, Italy}
\newcommand{\INFNFerrara}{Istituto Nazionale di Fisica Nucleare Sezione di Ferrara, I-44122 Ferrara, Italy}
\newcommand{\INFNFrascati}{Istituto Nazionale di Fisica Nucleare Laboratori Nazionali di Frascati, Frascati, Roma, Italy}
\newcommand{\INFNGenova}{Istituto Nazionale di Fisica Nucleare Sezione di Genova, 16146 Genova GE, Italy}
\newcommand{\INFNLecce}{Istituto Nazionale di Fisica Nucleare Sezione di Lecce, 73100 - Lecce, Italy}
\newcommand{\INFNMilanBicocca}{Istituto Nazionale di Fisica Nucleare Sezione di Milano Bicocca, 3 - I-20126 Milano, Italy}
\newcommand{\INFNMilano}{Istituto Nazionale di Fisica Nucleare Sezione di Milano, 20133 Milano, Italy}
\newcommand{\INFNNapoli}{Istituto Nazionale di Fisica Nucleare Sezione di Napoli, I-80126 Napoli, Italy}
\newcommand{\INFNPadova}{Istituto Nazionale di Fisica Nucleare Sezione di Padova, 35131 Padova, Italy}
\newcommand{\INFNPavia}{Istituto Nazionale di Fisica Nucleare Sezione di Pavia,  I-27100 Pavia, Italy}
\newcommand{\INFNPisa}{Istituto Nazionale di Fisica Nucleare Laboratori Nazionali di Pisa, Pisa PI, Italy}
\newcommand{\INFNRoma}{Istituto Nazionale di Fisica Nucleare Sezione di Roma, 00185 Roma RM, Italy}
\newcommand{\INFNRomavergata}{Istituto Nazionale di Fisica Nucleare Roma Tor Vergata , 00133 Roma RM, Italy}
\newcommand{\INFNSud}{Istituto Nazionale di Fisica Nucleare Laboratori Nazionali del Sud, 95123 Catania, Italy}
\newcommand{\Infntorino}{Istituto Nazionale di Fisica Nucleare, Sezione di Torino, Turin, Italy}
\newcommand{\Ingenieria}{Universidad Nacional de Ingenier\'ia, Lima 25, Per\'u}
\newcommand{\Insubria }{University of Insubria, Via Ravasi, 2, 21100 Varese VA, Italy}
\newcommand{\Iowa}{University of Iowa, Iowa City, IA 52242, USA}
\newcommand{\IowaState}{Iowa State University, Ames, Iowa 50011, USA}
\newcommand{\IPLyon}{Institut de Physique des 2 Infinis de Lyon, 69622 Villeurbanne, France}
\newcommand{\IPM}{Institute for Research in Fundamental Sciences, Tehran, Iran}
\newcommand{\IRLPPC}{Particle Physics and Cosmology International Research Laboratory, Chicago IL 60637, USA}
\newcommand{\ISTlisboa}{Instituto Superior T\'ecnico - IST, Universidade de Lisboa, 1049-001 Lisboa, Portugal}
\newcommand{\Ita}{Instituto Tecnol\'ogico de Aeron\'autica, Sao Jose dos Campos, Brazil}
\newcommand{\Iwate}{Iwate University, Morioka, Iwate 020-8551, Japan}
\newcommand{\Jacksonstate}{Jackson State University, Jackson, MS 39217, USA}
\newcommand{\Jawaharlal}{Jawaharlal Nehru University, New Delhi 110067, India}
\newcommand{\Jeonbuk}{Jeonbuk National University, Jeonrabuk-do 54896, South Korea}
\newcommand{\Jyvaskyla}{Jyv\"askyl\"a University, FI-40014 Jyv\"askyl\"a, Finland}
\newcommand{\Kansasstate}{Kansas State University, Manhattan, KS 66506, USA}
\newcommand{\Kavli}{Kavli Institute for the Physics and Mathematics of the Universe, Kashiwa, Chiba 277-8583, Japan}
\newcommand{\KEK}{High Energy Accelerator Research Organization (KEK), Ibaraki, 305-0801, Japan}
\newcommand{\KISTI}{Korea Institute of Science and Technology Information, Daejeon, 34141, South Korea}
\newcommand{\Kyiv}{Taras Shevchenko National University of Kyiv, 01601 Kyiv, Ukraine}
\newcommand{\Lancaster}{Lancaster University, Lancaster LA1 4YB, United Kingdom}
\newcommand{\LawrenceBerkeley}{Lawrence Berkeley National Laboratory, Berkeley, CA 94720, USA}
\newcommand{\LIP}{Laborat\'orio de Instrumenta{\c c}\~ao e F\'isica Experimental de Part\'iculas, 1649-003 Lisboa and 3004-516 Coimbra, Portugal}
\newcommand{\Liverpool}{University of Liverpool, L69 7ZE, Liverpool, United Kingdom}
\newcommand{\LosAlmos}{Los Alamos National Laboratory, Los Alamos, NM 87545, USA}
\newcommand{\Louisanastate}{Louisiana State University, Baton Rouge, LA 70803, USA}
\newcommand{\LpBordeaux}{Laboratoire de Physique des Deux Infinis Bordeaux - IN2P3, F-33175 Gradignan, Bordeaux, France }
\newcommand{\Lucknow}{University of Lucknow, Uttar Pradesh 226007, India}
\newcommand{\Mainz}{Johannes Gutenberg-Universit\"at Mainz, 55122 Mainz, Germany}
\newcommand{\Manchester}{University of Manchester, Manchester M13 9PL, United Kingdom}
\newcommand{\Massinsttech}{Massachusetts Institute of Technology, Cambridge, MA 02139, USA}
\newcommand{\Medellin}{University of Medell\'in, Medell\'in, 050026 Colombia }
\newcommand{\Michigan}{University of Michigan, Ann Arbor, MI 48109, USA}
\newcommand{\Michiganstate}{Michigan State University, East Lansing, MI 48824, USA}
\newcommand{\MilanoBicocca}{Universit\`a di Milano Bicocca , 20126 Milano, Italy}
\newcommand{\MilanoUniv}{Universit\`a degli Studi di Milano, I-20133 Milano, Italy}
\newcommand{\Minnduluth}{University of Minnesota Duluth, Duluth, MN 55812, USA}
\newcommand{\Minntwin}{University of Minnesota Twin Cities, Minneapolis, MN 55455, USA}
\newcommand{\Mississippi}{University of Mississippi, University, MS 38677 USA}
\newcommand{\napoli}{Universit\`a degli Studi di Napoli Federico II , 80138 Napoli NA, Italy}
\newcommand{\Nikhef}{Nikhef National Institute of Subatomic Physics, 1098 XG Amsterdam, Netherlands}
\newcommand{\Niser}{National Institute of Science Education and Research (NISER), Odisha 752050, India}
\newcommand{\Northdakota}{University of North Dakota, Grand Forks, ND 58202-8357, USA}
\newcommand{\Northernillinois}{Northern Illinois University, DeKalb, IL 60115, USA}
\newcommand{\Northwestern}{Northwestern University, Evanston, Il 60208, USA}
\newcommand{\NotreDame}{University of Notre Dame, Notre Dame, IN 46556, USA}
\newcommand{\NoviSad}{University of Novi Sad, 21102 Novi Sad, Serbia}
\newcommand{\Ohiostate}{Ohio State University, Columbus, OH 43210, USA}
\newcommand{\OregonState}{Oregon State University, Corvallis, OR 97331, USA}
\newcommand{\Oxford}{University of Oxford, Oxford, OX1 3RH, United Kingdom}
\newcommand{\PacificNorthwest}{Pacific Northwest National Laboratory, Richland, WA 99352, USA}
\newcommand{\Padova}{Universt\`a degli Studi di Padova, I-35131 Padova, Italy}
\newcommand{\Panjab}{Panjab University, Chandigarh, 160014, India}
\newcommand{\Parissaclay}{Universit\'e Paris-Saclay, CNRS/IN2P3, IJCLab, 91405 Orsay, France}
\newcommand{\Parisuniversite}{Universit\'e Paris Cit\'e, CNRS, Astroparticule et Cosmologie, Paris, France}
\newcommand{\Parma}{University of Parma,  43121 Parma PR, Italy}
\newcommand{\Pavia}{Universit\`a degli Studi di Pavia, 27100 Pavia PV, Italy}
\newcommand{\Penn}{University of Pennsylvania, Philadelphia, PA 19104, USA}
\newcommand{\PennState}{Pennsylvania State University, University Park, PA 16802, USA}
\newcommand{\PhysicalResearchLaboratory}{Physical Research Laboratory, Ahmedabad 380 009, India}
\newcommand{\Pisa}{Universit\`a di Pisa, I-56127 Pisa, Italy}
\newcommand{\Pitt}{University of Pittsburgh, Pittsburgh, PA 15260, USA}
\newcommand{\Pontificia}{Pontificia Universidad Cat\'olica del Per\'u, Lima, Per\'u}
\newcommand{\PuertoRico}{University of Puerto Rico, Mayaguez 00681, Puerto Rico, USA}
\newcommand{\Punjab}{Punjab Agricultural University, Ludhiana 141004, India}
\newcommand{\QMUL}{Queen Mary University of London, London E1 4NS, United Kingdom
}
\newcommand{\Radboud}{Radboud University, NL-6525 AJ Nijmegen, Netherlands}
\newcommand{\Rice}{Rice University, Houston, TX 77005, USA}
\newcommand{\Rochester}{University of Rochester, Rochester, NY 14627, USA}
\newcommand{\Royalholloway}{Royal Holloway College London, London, TW20 0EX, United Kingdom}
\newcommand{\Rutgers}{Rutgers University, Piscataway, NJ, 08854, USA}
\newcommand{\Rutherford}{STFC Rutherford Appleton Laboratory, Didcot OX11 0QX, United Kingdom}
\newcommand{\Salento}{Universit\`a del Salento, 73100 Lecce, Italy}
\newcommand{\santamarta}{Universidad del Magdalena, Santa Marta - Colombia}
\newcommand{\Sapienza}{Sapienza University of Rome, 00185 Roma RM, Italy}
\newcommand{\SergioArboleda}{Universidad Sergio Arboleda, 11022 Bogot\'a, Colombia}
\newcommand{\Sheffield}{University of Sheffield, Sheffield S3 7RH, United Kingdom}
\newcommand{\SLAC}{SLAC National Accelerator Laboratory, Menlo Park, CA 94025, USA}
\newcommand{\Southcarolina}{University of South Carolina, Columbia, SC 29208, USA}
\newcommand{\SouthDakotaSchool}{South Dakota School of Mines and Technology, Rapid City, SD 57701, USA}
\newcommand{\SouthDakotaState}{South Dakota State University, Brookings, SD 57007, USA}
\newcommand{\StonyBrook}{Stony Brook University, SUNY, Stony Brook, NY 11794, USA}
\newcommand{\SURF}{Sanford Underground Research Facility, Lead, SD, 57754, USA}
\newcommand{\Sussex}{University of Sussex, Brighton, BN1 9RH, United Kingdom}
\newcommand{\Syracuse}{Syracuse University, Syracuse, NY 13244, USA}
\newcommand{\Tecnologica }{Universidade Tecnol\'ogica Federal do Paran\'a, Curitiba, Brazil}
\newcommand{\TelAviv}{Tel Aviv University, Tel Aviv-Yafo, Israel}
\newcommand{\TexasAMcollege}{Texas A\&M University, College Station, Texas 77840, USA}
\newcommand{\TexasAMcorpuscristi}{Texas A\&M University - Corpus Christi, Corpus Christi, TX 78412, USA}
\newcommand{\TexasArlington}{University of Texas at Arlington, Arlington, TX 76019, USA}
\newcommand{\Texasaustin}{University of Texas at Austin, Austin, TX 78712, USA}
\newcommand{\Toronto}{University of Toronto, Toronto, Ontario M5S 1A1, Canada}
\newcommand{\Tufts}{Tufts University, Medford, MA 02155, USA}
\newcommand{\Unifesp}{Universidade Federal de S\~ao Paulo, 09913-030, S\~ao Paulo, Brazil}
\newcommand{\UNIST}{Ulsan National Institute of Science and Technology, Ulsan 689-798, South Korea}
\newcommand{\UniversityCollegeLondon}{University College London, London, WC1E 6BT, United Kingdom}
\newcommand{\univkansas}{University of Kansas, Lawrence, KS 66045, USA}
\newcommand{\UNMSM}{Universidad Nacional Mayor de San Marcos, Lima, Peru}
\newcommand{\ValleyCity}{Valley City State University, Valley City, ND 58072, USA}
\newcommand{\Vigo}{University of Vigo, E- 36310 Vigo Spain}
\newcommand{\VirginiaTech}{Virginia Tech, Blacksburg, VA 24060, USA}
\newcommand{\Warsaw}{University of Warsaw, 02-093 Warsaw, Poland}
\newcommand{\Warwick}{University of Warwick, Coventry CV4 7AL, United Kingdom}
\newcommand{\Wellesley}{Wellesley College, Wellesley, MA 02481, USA}
\newcommand{\Wichita}{Wichita State University, Wichita, KS 67260, USA}
\newcommand{\WilliamMary}{William and Mary, Williamsburg, VA 23187, USA}
\newcommand{\Wisconsin}{University of Wisconsin Madison, Madison, WI 53706, USA}
\newcommand{\Yale}{Yale University, New Haven, CT 06520, USA}
\newcommand{\Yerevan}{Yerevan Institute for Theoretical Physics and Modeling, Yerevan 0036, Armenia}
\newcommand{\York}{York University, Toronto M3J 1P3, Canada}
\begin{document}


\title{Cosmic ray muon seasonal variations with the ProtoDUNE-SP Detector}

\affiliation{\Albanysuny}
\affiliation{\Almaty}
\affiliation{\Amsterdam}
\affiliation{\Antalya}
\affiliation{\Antananarivo}
\affiliation{\Antioquia}
\affiliation{\AntonioNarino}
\affiliation{\Argonne}
\affiliation{\Arizona}
\affiliation{\Asuncion}
\affiliation{\Athens}
\affiliation{\Atlantico}
\affiliation{\Augustana}
\affiliation{\Bern}
\affiliation{\Beykent}
\affiliation{\Birmingham}
\affiliation{\BolognaUniversity}
\affiliation{\Boston}
\affiliation{\Bristol}
\affiliation{\Brookhaven}
\affiliation{\Bucharest}
\affiliation{\CalBerkeley}
\affiliation{\CalDavis}
\affiliation{\CalIrvine}
\affiliation{\CalLosangeles}
\affiliation{\CalRiverside}
\affiliation{\CalSantabarbara}
\affiliation{\Caltech}
\affiliation{\Cambridge}
\affiliation{\Campinas}
\affiliation{\CataniaUniversitadi}
\affiliation{\Catolica}
\affiliation{\CBPF}
\affiliation{\CEASaclay}
\affiliation{\CERN}
\affiliation{\Charles}
\affiliation{\Chicago}
\affiliation{\ChungAng}
\affiliation{\CIEMAT}
\affiliation{\Cincinnati}
\affiliation{\Cinvestav}
\affiliation{\Colima}
\affiliation{\ColoradoBoulder}
\affiliation{\ColoradoState}
\affiliation{\Columbia}
\affiliation{\conida}
\affiliation{\Cti}
\affiliation{\CUSB}
\affiliation{\CzechAcademyofSciences}
\affiliation{\CzechTechnical}
\affiliation{\DannecyleVieux}
\affiliation{\Daresbury}
\affiliation{\Dordt}
\affiliation{\Drexel}
\affiliation{\Duke}
\affiliation{\Edinburgh}
\affiliation{\EIA}
\affiliation{\Eotvos}
\affiliation{\erciyes}
\affiliation{\FCULport}
\affiliation{\FederaldeAlfenas}
\affiliation{\FederaldeGoias}
\affiliation{\FederaldoABC}
\affiliation{\FederaldoRio}
\affiliation{\Fermi}
\affiliation{\Ferrarauniv}
\affiliation{\Florida}
\affiliation{\Floridastate}
\affiliation{\Fluminense}
\affiliation{\Genova}
\affiliation{\Georgian}
\affiliation{\Granada}
\affiliation{\GranSasso}
\affiliation{\GranSassoLab}
\affiliation{\Grenoble}
\affiliation{\Guanajuato}
\affiliation{\Harish}
\affiliation{\Hawaii}
\affiliation{\hkust}
\affiliation{\Houston}
\affiliation{\Hyderabad}
\affiliation{\Idaho}
\affiliation{\IFIC}
\affiliation{\IGFAE}
\affiliation{\ihep}
\affiliation{\Iitk}
\affiliation{\Illinoisinstitute}
\affiliation{\Imperial}
\affiliation{\IndGuwahati}
\affiliation{\IndHyderabad}
\affiliation{\Indiana}
\affiliation{\INFNBologna}
\affiliation{\INFNCatania}
\affiliation{\INFNFerrara}
\affiliation{\INFNFrascati}
\affiliation{\INFNGenova}
\affiliation{\INFNLecce}
\affiliation{\INFNMilanBicocca}
\affiliation{\INFNMilano}
\affiliation{\INFNNapoli}
\affiliation{\INFNPadova}
\affiliation{\INFNPavia}
\affiliation{\INFNPisa}
\affiliation{\INFNRoma}
\affiliation{\INFNRomavergata}
\affiliation{\INFNSud}
\affiliation{\Infntorino}
\affiliation{\Ingenieria}
\affiliation{\Insubria }
\affiliation{\Iowa}
\affiliation{\IowaState}
\affiliation{\IPLyon}
\affiliation{\IPM}
\affiliation{\IRLPPC}
\affiliation{\ISTlisboa}
\affiliation{\Ita}
\affiliation{\Iwate}
\affiliation{\Jacksonstate}
\affiliation{\Jawaharlal}
\affiliation{\Jeonbuk}
\affiliation{\Jyvaskyla}
\affiliation{\Kansasstate}
\affiliation{\Kavli}
\affiliation{\KEK}
\affiliation{\KISTI}
\affiliation{\Kyiv}
\affiliation{\Lancaster}
\affiliation{\LawrenceBerkeley}
\affiliation{\LIP}
\affiliation{\Liverpool}
\affiliation{\LosAlmos}
\affiliation{\Louisanastate}
\affiliation{\LpBordeaux}
\affiliation{\Lucknow}
\affiliation{\Mainz}
\affiliation{\Manchester}
\affiliation{\Massinsttech}
\affiliation{\Medellin}
\affiliation{\Michigan}
\affiliation{\Michiganstate}
\affiliation{\MilanoBicocca}
\affiliation{\MilanoUniv}
\affiliation{\Minnduluth}
\affiliation{\Minntwin}
\affiliation{\Mississippi}
\affiliation{\napoli}
\affiliation{\Nikhef}
\affiliation{\Niser}
\affiliation{\Northdakota}
\affiliation{\Northernillinois}
\affiliation{\Northwestern}
\affiliation{\NotreDame}
\affiliation{\NoviSad}
\affiliation{\Ohiostate}
\affiliation{\OregonState}
\affiliation{\Oxford}
\affiliation{\PacificNorthwest}
\affiliation{\Padova}
\affiliation{\Panjab}
\affiliation{\Parissaclay}
\affiliation{\Parisuniversite}
\affiliation{\Parma}
\affiliation{\Pavia}
\affiliation{\Penn}
\affiliation{\PennState}
\affiliation{\PhysicalResearchLaboratory}
\affiliation{\Pisa}
\affiliation{\Pitt}
\affiliation{\Pontificia}
\affiliation{\PuertoRico}
\affiliation{\Punjab}
\affiliation{\QMUL}
\affiliation{\Radboud}
\affiliation{\Rice}
\affiliation{\Rochester}
\affiliation{\Royalholloway}
\affiliation{\Rutgers}
\affiliation{\Rutherford}
\affiliation{\Salento}
\affiliation{\santamarta}
\affiliation{\Sapienza}
\affiliation{\SergioArboleda}
\affiliation{\Sheffield}
\affiliation{\SLAC}
\affiliation{\Southcarolina}
\affiliation{\SouthDakotaSchool}
\affiliation{\SouthDakotaState}
\affiliation{\StonyBrook}
\affiliation{\SURF}
\affiliation{\Sussex}
\affiliation{\Syracuse}
\affiliation{\Tecnologica }
\affiliation{\TelAviv}
\affiliation{\TexasAMcollege}
\affiliation{\TexasAMcorpuscristi}
\affiliation{\TexasArlington}
\affiliation{\Texasaustin}
\affiliation{\Toronto}
\affiliation{\Tufts}
\affiliation{\Unifesp}
\affiliation{\UNIST}
\affiliation{\UniversityCollegeLondon}
\affiliation{\univkansas}
\affiliation{\UNMSM}
\affiliation{\ValleyCity}
\affiliation{\Vigo}
\affiliation{\VirginiaTech}
\affiliation{\Warsaw}
\affiliation{\Warwick}
\affiliation{\Wellesley}
\affiliation{\Wichita}
\affiliation{\WilliamMary}
\affiliation{\Wisconsin}
\affiliation{\Yale}
\affiliation{\Yerevan}
\affiliation{\York}

\author{S.~Abbaslu} \affiliation{\IPM}
\author{A.~Abed Abud} \affiliation{\CERN}
\author{R.~Acciarri} \affiliation{\Fermi}
\author{L.~P.~Accorsi} \affiliation{\Tecnologica }
\author{M.~A.~Acero} \affiliation{\Atlantico}
\author{M.~R.~Adames} \affiliation{\Tecnologica }
\author{G.~Adamov} \affiliation{\Georgian}
\author{M.~Adamowski} \affiliation{\Fermi}
\author{C.~Adriano} \affiliation{\Campinas}
\author{F.~Akbar} \affiliation{\Rochester}
% ...

\newif\ifmarkup
\newcommand{\added}[1]{\ifmarkup{\color{blue}#1}\else#1\fi}
\newcommand{\deleted}[1]{\ifmarkup{\color{red}\sout{#1}}\else\fi}
 
\begin{abstract}
We present the measurement of the cosmic ray muon seasonal variation using single-muon events with the ProtoDUNE single-phase detector located at the CERN Neutrino Platform, using data collected from December 2018 to July 2020. Cosmic ray muon runs recorded during this period are analyzed to determine the muon rate as a function of time. The results show a seasonal modulation of the muon rate, with a significance of $4\,\sigma$ relative to the null hypothesis of no seasonal variation. The observed muon rate reaches a maximum during the winter, consistent with measurements from other \deleted{experiments}\added{surface experiments with a low muon-energy threshold}. This measurement is possible due to the overall excellent performance of this detector technology and serves as a clear demonstration that such a detector can stably measure small, time-dependent physical signals such as seasonal variations in cosmic muon rates over long timescales.
\end{abstract}
 
\maketitle
 
\newpage
 
\section{Introduction}
 
The Deep Underground Neutrino Experiment (DUNE) \cite{abi2020deep} will be hosted by the Fermi National Accelerator Laboratory (Fermilab) near Chicago, Illinois, USA. It will utilize an intense neutrino beam produced at Fermilab and directed through the Earth over a baseline of 1300~km to the Sanford Underground Research Facility (SURF), where the far detector will be installed approximately 1.5~km underground. The far detector will consist of four modules containing a total of 70~kt of liquid argon. Its primary physics goal is the study of neutrino oscillations through the disappearance of muon neutrinos and the appearance of electron neutrinos.
 
The DUNE collaboration constructed and operated a horizontal-drift prototype detector, ProtoDUNE-SP, as a prototype of the DUNE far detector~\cite{DUNE:2020cqd}. The detector consists of an $8.5 \times 8.5 \times 7.9$~m$^3$ cryostat (dimensions in the $x$, $z$, and $y$ directions, respectively), containing approximately 800~t of liquid argon. We define the coordinate system such that the $x$-axis is along the electron drift direction, the $y$-axis is vertical (positive upward), and the $z$-axis is parallel to the beam direction. It comprises two drift volumes, each with a drift distance of 3.6~m; under the nominal electric field of 500~V/cm, this corresponds to a maximum electron drift time of 2.25~ms. The detector was assembled and tested at the CERN Neutrino Platform \cite{Pietropaolo_2017, DUNE:2017pqt} and was operated from December 2018 to July 2020 \cite{Adams:2020tqx}. During this period, it collected beam and cosmic ray (CR) data samples that were used to demonstrate the performance of the single-phase liquid argon time projection chamber technology for the DUNE far detector. A detailed characterization of the ProtoDUNE-SP liquid argon time projection chamber (TPC) performance is reported in~\cite{DUNE:2021hwx}. This includes measurements of the electronic noise and gain, $dE/dx$ calibrations using muons, protons, pions, and electrons, drift-electron lifetime studies, and evaluations of the photon-detector noise, signal sensitivity, and time resolution. Results on the performance of track, shower, and event reconstruction can be found in \cite{DUNE:2020cqd}. 
 
Cosmic ray muons provide a stable reference signal to study the performance and time stability of detectors, calibrate their responses, and evaluate reconstruction algorithms. In addition, the analysis of CR muons permits the measurements of various physical phenomena \cite{ICARUS:2004michel,Arunbabu:2017imc,PhysRevD.99.122004}\added{ (see Refs.~\cite{Gaisser2016,PDG2024} for reviews of the atmospheric muon flux)}.
 
The flux of CR muons observed in underground and surface detectors\deleted{, shown in Table~I,} exhibits a seasonal variation that is correlated with atmospheric temperature\added{~\cite{Duperier1949,Barrett1952,PhysRevD.99.122004,Grashorn_2010}}. This effect is well understood as a consequence of temperature-driven density variations in the atmosphere\added{~\cite{Gaisser2016,PDG2024}}, which act as the target medium for primary cosmic-ray interactions. Mesons ($K^\pm$, $\pi^\pm$) produced in these interactions may subsequently either interact or decay, generating muons and muon neutrinos. Specifically, during summer, the increased atmospheric temperature causes the atmosphere to expand, reducing the probability of secondary-meson interactions. Therefore, a leading pion produced in the primary interaction is more likely to decay during summer~\cite{PhysRevD.99.122004,Grashorn_2010}. In winter, it is instead more likely to undergo a secondary hadronic interaction.
 
For underground detectors with a muon energy threshold of 10 GeV or greater, a maximum rate of single-muons is expected in the summer. At the lower muon energies of a few GeV relevant for surface detectors, seasonal differences in muon decay in flight can also become significant. In particular, during summer, these muons are typically produced at higher altitudes than in winter, increasing the probability that they decay in flight. As a result, the phase of the observed modulation \deleted{may depend}\added{depends} on the detector’s effective energy threshold.
In addition, the higher altitude muon production in the summer results in muon showers spreading out more by the time they reach the detectors, yielding fewer of them being identified as multi-muons and thus a lower multi-muon rate\added{~\cite{NOvA2026,Gaisser2021}}. 
 
In this paper, we present the measurement of the CR seasonal variation with the ProtoDUNE-SP detector operated as a surface detector. The analysis is straightforward and is designed to be robust. Section~II describes the data sample and the run-selection criteria used to ensure high-quality data runs and stable detector conditions across the full dataset, thereby maintaining consistent reconstruction over time. Section~III summarizes the reconstruction procedure and the selection of a high-purity CR muon sample. Section~IV presents the measurement of the muon rate and the associated systematic uncertainties. In Sec.~V, we discuss the observed seasonal modulation, its interpretation in terms of atmospheric effects, and comparisons with the simulation. Conclusions are presented in Sec.~VI.
 
\section{Data sets}
 
The ProtoDUNE-SP detector collected data for two years, from December 2018 to July 2020. During this period, data were collected using three sets of triggers: beam-related to select events from the CERN Super Proton Synchrotron (SPS), cosmic triggers recording the activity in a fixed 3 ms time window initiated by the passage of a charged particle from the top or the sides of the detector, 
and random triggers in which all detector activity is recorded in a random 3 ms time window with no beam present. In this analysis, we utilized random triggers in order to have an unbiased sample of CR muon events.  
 
In ProtoDUNE, a ``run'' corresponds to a continuous data-taking period defined by stable detector and DAQ configurations. The duration of a run is not fixed and typically ranges from about 30 minutes to a few hours. \deleted{For this analysis, one representative CR run per day was selected to ensure uniform temporal sampling.} We select cosmic runs recorded under nominal, stable, and similar detector conditions as far as the cathode high voltage, the number of dead or unsynchronized channels, and the argon purity\added{, with no special data-taking tests performed}. \deleted{In addition, we require that no special data-taking tests were performed during these runs.} \added{At most one run per day is used.} In total, \added{only} twelve CR runs, comprising 373,980 events, satisfied all of \deleted{our selection} \added{these} criteria and are utilized in this measurement. 
 
Simulated data are employed to cross-check the reconstruction performance and to confirm the absence of significant pathologies, in particular broken or badly reconstructed tracks, as discussed in the following section. For the simulated events, the CORSIKA \cite{Heck:1998vt} event generator\added{ (version 7.7400)} is utilized, with the detector description in Geant4 \cite{church2014larsoftsoftwarepackageliquid,mp3}\added{ (version v4\_10\_6\_p01)}. Details of the event reconstruction using both real and simulated data samples can be found in Ref. \cite{DUNE:2017pqt}.
 
\added{The CORSIKA showers were normalized to a power-law primary cosmic-ray flux with spectral index $-2.7$. The primary particles consisted of protons and helium, nitrogen, magnesium, and iron nuclei, with species-dependent minimum energies of 1.3, 5.2, 18.2, 31.2, and 72.8~GeV, respectively, and a maximum energy of 100~TeV. Zenith angles from $0^\circ$ to $90^\circ$ and the full azimuthal range were simulated. QGSJET~\cite{Ostapchenko2011} was used to model high-energy hadronic interactions, with an energy threshold of 0.05~GeV. The observation level, defined as the height above sea level where the properties and secondary particles of an extensive air shower are recorded, was set to 455~m, while an additional sample was generated at 1100~m for the seasonal study. Approximately $10^8$ air showers were generated.}
 
\added{To investigate the presence of multiple muons from the same air shower, muon bundles were defined by $\cos\theta>0.992$ and a time separation $\Delta t<100$~ns. In the 970 simulated readout windows considered for this study, the mean number of muons per event belonging to such bundles was 0.004 at truth level.}
 
\section{Event Reconstruction and Selection}
 
The ProtoDUNE-SP physics object and event reconstruction is described in detail in \cite{DUNE:2020cqd}. For this analysis, the reconstructed track candidates are used to define a high-purity CR muon sample. To achieve this, we define a set of CR muon selection criteria in order to reject delta rays and remove short, unphysical tracks from possible reconstruction failures and/or detector related effects. First, we restrict the analysis to the positive X section of the detector (X$>0$, beam right) to mitigate differences in track reconstruction between the positive and negative X sides. These differences are associated with the presence of Electron Diverters \cite{DUNE:2020cqd}. Next, we select tracks with a length greater than 15 cm, removing more than 95$\%$ of delta rays, ensuring that the selected muon track sample is of high purity. 
 
Then, we devised a set of geometrical, fiducial volume criteria
involving the starting and ending points of the reconstructed muon tracks, to further reduce unphysical tracks that cluster around detector boundaries close to the anode plane assemblies (APA), and to select tracks in regions of the detector that are not sensitive to a space-charge effect (SCE) \cite{DUNE:2020cqd} due to the detector operating on the surface. \deleted{Finally, we introduce an additional criterion that removes parallel tracks with a very small distance of minimum approach, coming from segmented muon tracks. This rejects only $2\%$ of the muon tracks surviving all previous criteria.} \added{Finally, to avoid double-counting muons that are falsely reconstructed as two separate track segments, we reject pairs of tracks satisfying both $|\cos\theta|>0.96$, where $\theta$ is the angle between their directions, and $\Delta S<12$~cm, where $\Delta S$ is the distance of closest approach between the tracks. Such pairs are consistent with the reconstruction of a single muon as two nearly collinear track segments. This requirement removes approximately $2\%$ of the tracks remaining after all preceding selection criteria.}
 
Following \cite{DUNE:2022wlc}, we define track completeness as the fraction of simulated truth particle hits shared with the reconstructed track, and track purity as the fraction of reconstructed track hits shared with the corresponding simulated truth particle. The track selection criteria described above improve track completeness and purity. Specifically, after these criteria are applied, 63\% of tracks exhibit completeness greater than 80\%, compared to 36\% before, and 97\% have a purity greater than 80\% compared to 89\% before, with an overall reconstruction efficiency greater than 96\%. 
 
We note that the limited size of the ProtoDUNE-SP detector makes the probability of observing multiple, nearly parallel muons originating from the same air shower within its fiducial volume negligible. \added{The fiducial volume is approximately $3.6\times 6.0\times 6.4$~m$^3\approx 138$~m$^3$, corresponding to a horizontal projected area of about 23~m$^2$.} Detailed simulation studies yield \deleted{zero such events given the available statistics}\added{a negligible number of such events (Sec.~II)}; thus, we do not report on multiple-muon events.
 
\section{Measurement of the muon rate}
 
\begin{figure}[htbp]
    \includegraphics[width=0.55\textwidth]{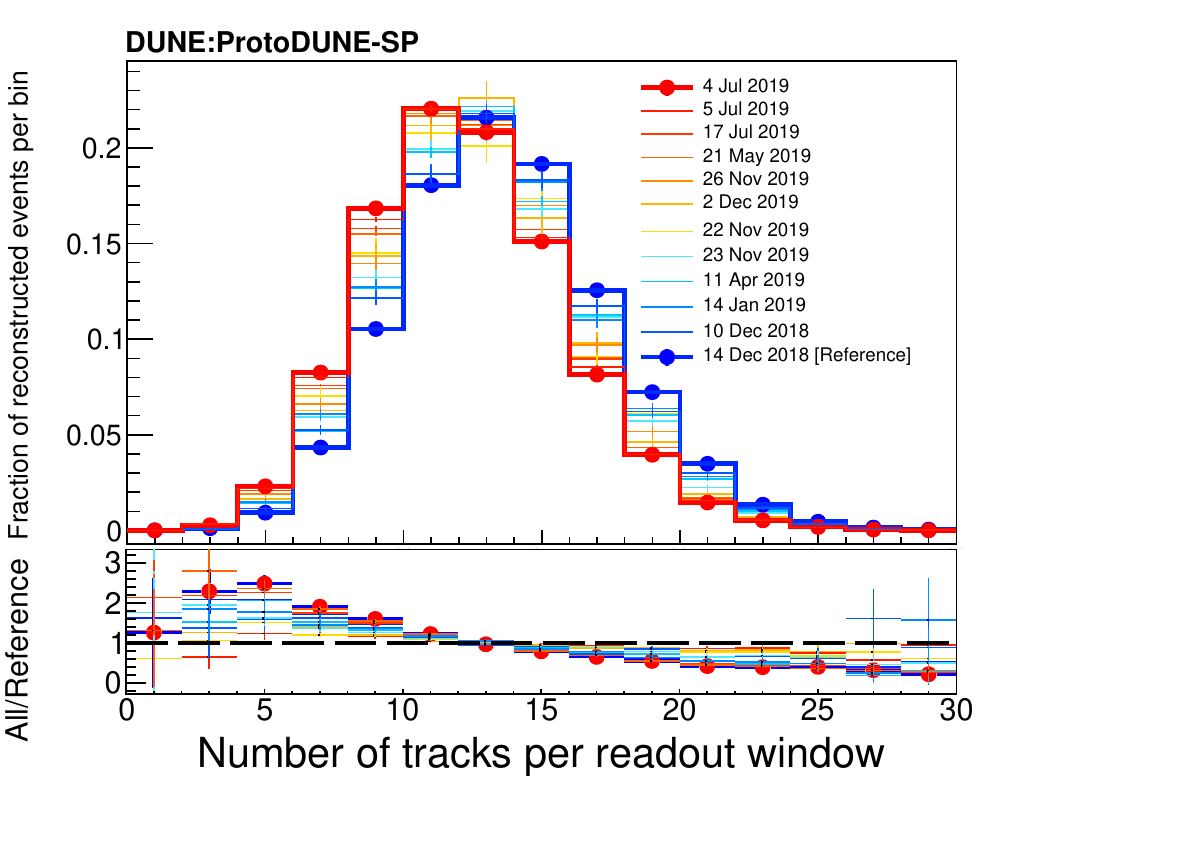}
    \caption{
        \justifying
    \deleted{The CR muon track multiplicity for the different runs under consideration. The track multiplicity varies across different CR runs, with winter runs like 14 Dec 2018 (dark blue) exhibiting higher peaks compared to summer runs like 4 Jul 2019 (red).} \added{Distribution of the number of reconstructed CR muon tracks per readout window for the different runs under consideration. Winter runs like 14 Dec 2018 (dark blue) peak at larger numbers of tracks than summer runs like 4 Jul 2019 (red).} The bottom panel illustrates the ratio of each CR run to a reference high-statistics CR run.}
   \label{fig:cosmic3}
\end{figure}

We estimate the CR muon rate for each run from the mean number of CR muons per triggered event with a readout time window of 3 ms. Figure \ref{fig:cosmic3} shows the distribution of the number of CR muons per readout window for each of the runs we considered. This distribution exhibits a clear variation as a function of time, which is interpreted as a seasonal variation and discussed in the next section. 
 
Since this measurement does not rely on simulation, and the seasonal variation of the CR muon rate is determined solely from comparisons of data collected during different running periods, only experimental systematic uncertainties are considered. Therefore, the only source of systematic uncertainty on the analysis comes from possible time-dependent variations in the detector performance that could affect track reconstruction, and hence the measured muon rate, which also follows a correlated seasonal pattern. Among these sources, the main ones are the effect of the time-varying argon purity \cite{DUNE:2020cqd}, and the variation of the SCE effect \cite{DUNE:2020cqd} due to the detector operating on the surface. 
 
To estimate the former, we generate and reconstruct simulated cosmic muon samples with varying purity in a similar magnitude as was measured by argon purity monitors during data-taking \cite{DUNE:2020cqd}. Then, analyzing these simulated data as the actual ones, we estimate the effect this has on the cosmic muon rate, which is found to be negligible.
 
For the SCE effect, we utilize a data-driven approach, taking into account the fact that the effect is most prominent in certain regions of the detector. For this, we estimate and compare the spatial track distortions for each of the considered runs. First, we examine if the measured spatial track distortions exhibit a seasonal variation effect that could affect our measurement, and we observe no such correlation. Next, we estimate the maximum spatial track distortion difference between the runs under consideration, which is found to be of the order of 8\%. Then, we generate simulated samples to estimate what effect this difference has on the measured cosmic muon rate. The estimated variation of the cosmic muon rate resulting from such SCE variations is 2.3$\%$, and it is utilized as an uncorrelated run-by-run systematic uncertainty. We note that systematic uncertainty is by far the dominant factor in this measurement.
 
\begin{figure}[htbp]
    \captionsetup{
        format=plain,
        singlelinecheck=false,
        width=0.48\textwidth
    }
    \includegraphics[width= 0.55\textwidth]{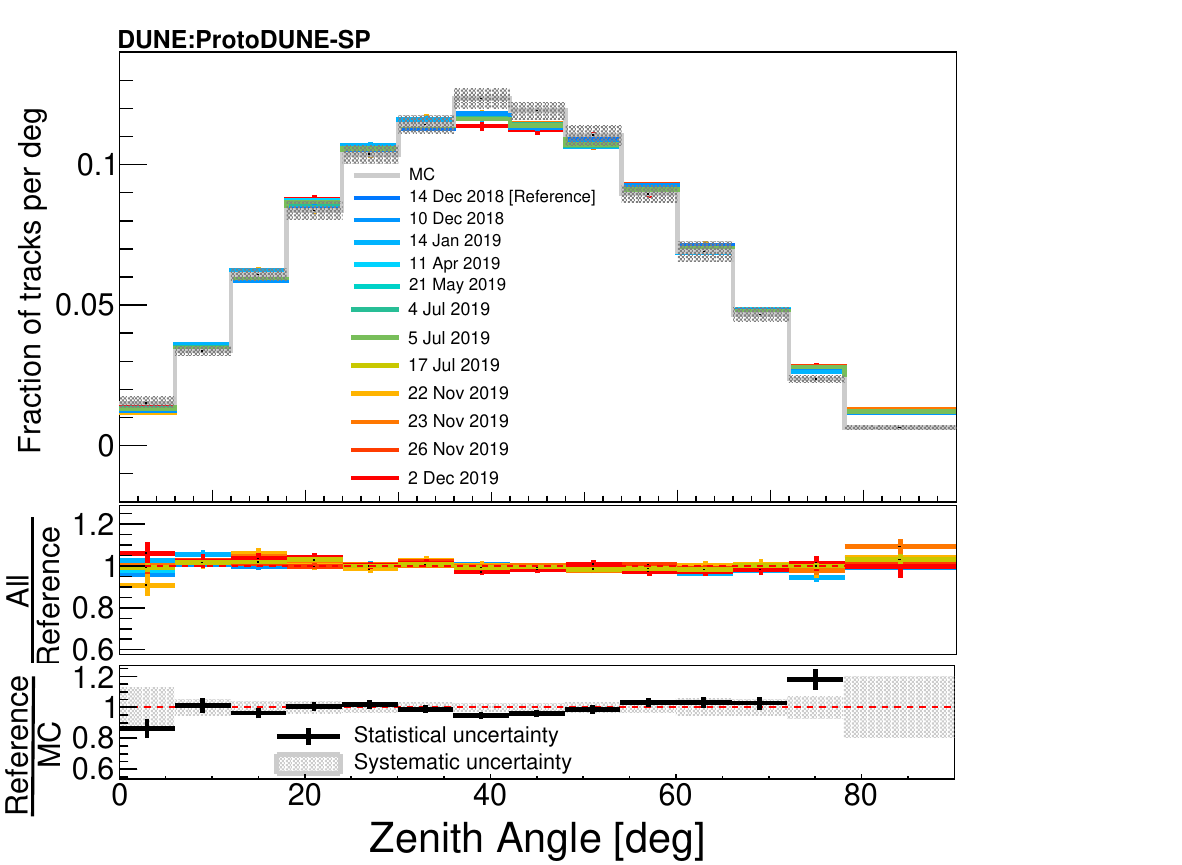}
    \includegraphics[width= 0.55\textwidth]{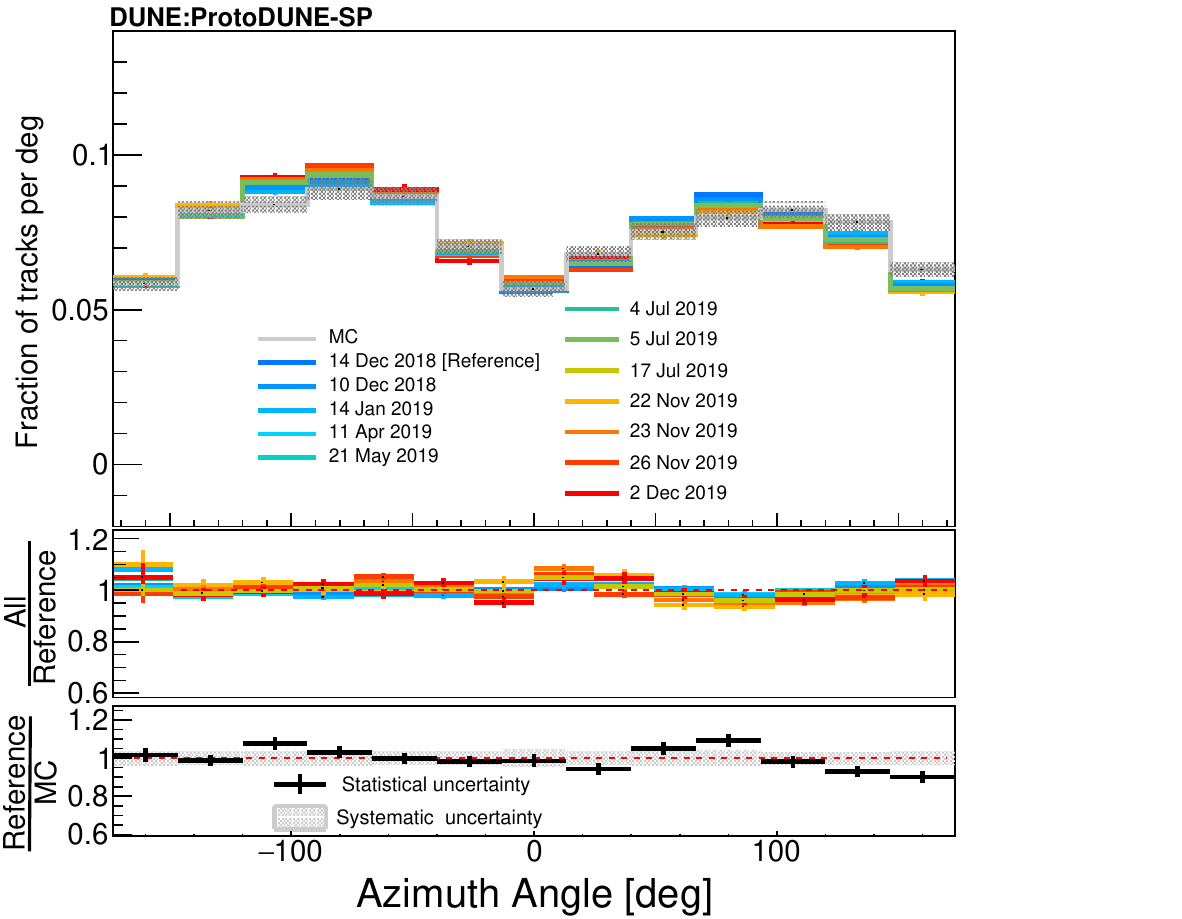}
    \caption{
            \justifying
    The CR muon zenith angle (top) and azimuth angle (bottom) for each of the utilized CR runs. The middle panel illustrates the ratio of each CR run to a reference high-statistics CR run. The bottom panel illustrates the ratio of data (reference) to simulation with statistical black error bars and systematic gray band uncertainties shown. }
   \label{fig:cosmic1}
\end{figure}

\added{We also considered a possible seasonal dependence of the track sample purity, through other atmospheric particles reaching the detector or secondaries produced in the surrounding material. After all selections, electrons and photons make up 1.6\% of the tracks and muons 96.8\%. Even a conservative 4\% contamination varying seasonally by as much as the muon rate itself ($\sim$10\%) would change the rate by less than 0.4\%, well below the 2.3\% systematic uncertainty. Contaminants produced by muons scale with the muon flux and do not change the relative amplitude. The result is also stable against stricter track-length requirements ($>100$~cm and $>200$~cm), which strongly suppress such contamination, and the angular distributions are the same in winter and summer (Fig.~\ref{fig:cosmic1}).}
 
In order to examine the quality and time stability of the selected CR muons, we perform comparisons of their characteristics between data and simulation, and among the different data runs. In Fig. \ref{fig:cosmic1}, we show the distributions of the CR muon zenith and azimuth angles for the different data runs and for real and simulated data. \added{The double-peaked azimuth shape is already present at generator level for muons entering the detector and is therefore an acceptance effect of the detector geometry. Winter and summer runs show no significant azimuth difference, and the per-run ratios to the reference are compatible with unity within uncertainties. }
 
\section{Results and interpretation}
 
In order to examine whether a seasonal modulation is present, we study the cosmic-muon rate as a function of time and model it with the sinusoidal form
 
\begin{equation}
R_\mu(t)=\langle R_\mu\rangle + \Delta R_\mu \cos\left[\frac{2\pi}{T}(t-t_0)\right],
\label{eq:absolute}
\end{equation}
where $\langle R_\mu\rangle$ is the mean muon rate, $\Delta R_\mu$ is the modulation amplitude, $T$ is the period, fixed to 365~days, and $t_0$ is the phase corresponding to the day at which the muon rate reaches its maximum. Here, $t$ denotes the number of days elapsed since January~1st of each year.
 
\begin{figure}[htbp]
    \captionsetup{
        format=plain,
        singlelinecheck=false,
        width=0.48\textwidth
    }
     \includegraphics[width=0.5\textwidth]{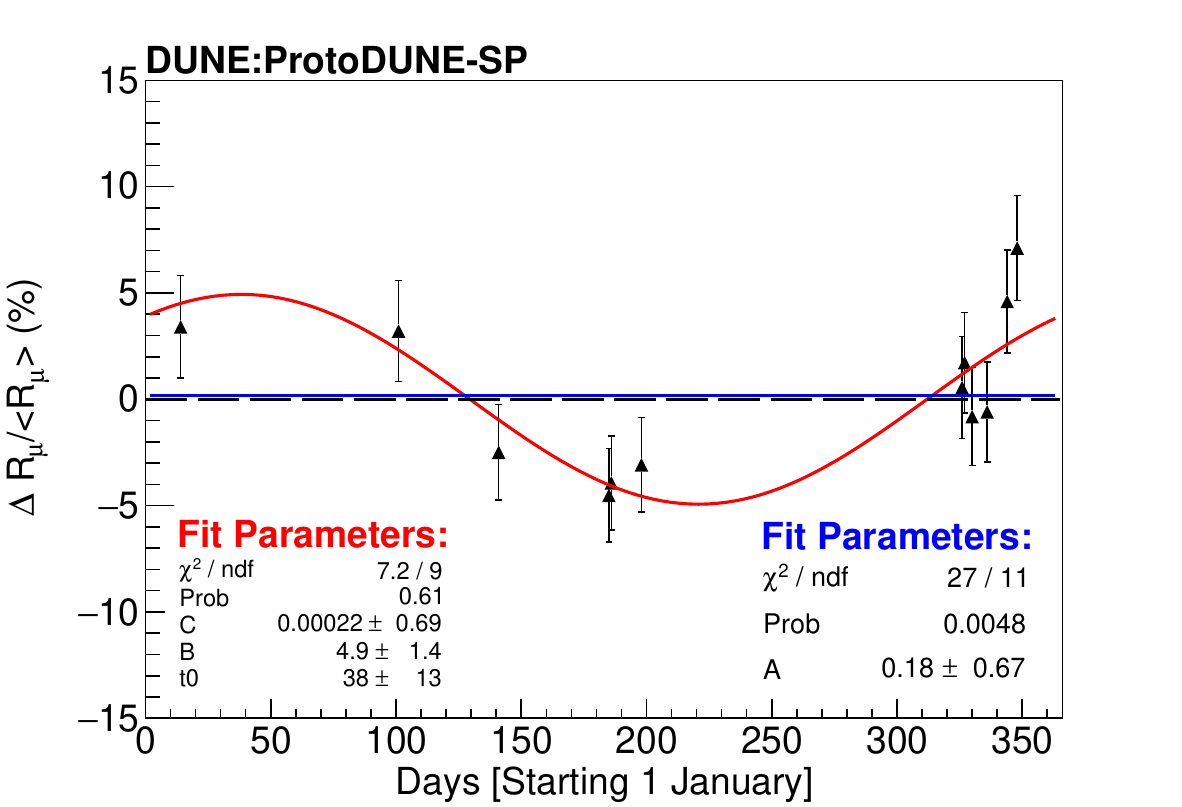}
    \caption{
          \justifying
    The percent change of CR muon rate as a function of time (day of year),  fitted under a seasonal variation hypothesis (red line) and one where the seasonal muon rate is constant (blue line). Cosmic ray runs are arranged according to the day they were taken within the year,
starting from January 1st. The error bars represent the quadratic sum of statistical and systematic uncertainties. The $\Delta R_{\mu}$ is the difference from the mean rate $< R_{\mu}>$.}
 \label{seasonal}
 \end{figure}

For visualization, the measured muon rate is subsequently normalized to its mean value. Then, the seasonal variation hypothesis is tested by fitting the relative rate variation, as shown in Fig.~\ref{seasonal}, with the functional form below, resulting directly from eq.~\ref{eq:absolute}:
 
\begin{equation}
\frac{\Delta R_\mu}{\langle R_\mu \rangle}(t)
=
C + B\cos\left[\frac{2\pi}{T}(t-t_0)\right],
\label{eq3}
\end{equation}
where $C$ is a constant with an expected value of zero, and $B$ is the relative modulation amplitude expressed in percent. The parameters $C$, $B$, and $t_0$ are treated as free parameters in the fit. \added{The single-sinusoid form is an approximation. The actual variation need not be symmetric between spring and autumn and is affected by shorter-term atmospheric variations. With denser time sampling, more detailed methods based on the measured effective atmospheric temperature would be preferable (see e.g.\ Refs.~\cite{OPERA:2018jif,Agostini_2019}).}
 
To estimate the significance of the measurement, we perform two distinct fits: one using a horizontal line to model no seasonal variation, and the other using the model described above, as shown in Fig.~\ref{seasonal}. We obtain $\Delta \chi^2 \approx19.8$ for a difference in degrees of freedom of 2, corresponding to a significance of approximately $4\sigma$ in favor of the seasonal variation hypothesis over the no-variation hypothesis. \added{The significance was verified using $10^7$ pseudo-experiments, giving a $p$-value of $5.3\times 10^{-5}$ ($4.04\sigma$).}
 
The fitted amplitude is $(4.9 \pm 1.4)\%$ and the maximum muon rate is taking place in winter at $38 \pm 13$ days, which is in agreement with the majority of measurements from other surface experiments sampling mostly multi-muon events ~\cite{Kokoulin:2015xwv,NOvA:2021lsm,PhysRevD.99.122004}\deleted{ as shown in Table~I}. \added{A winter maximum is observed by surface experiments with low energy thresholds~\cite{Arunbabu:2017imc,Kokoulin:2015xwv,NOvA:2021lsm}, whereas single muons in deep underground detectors peak in summer~\cite{MINOS:2015ikk,Ambrosio_2003,collaboration_2016,article,OPERA:2018jif,Agostini_2019}. For overviews, see Refs.~\cite{Verpoest2024,OPERA:2018jif}.} In the case of ProtoDUNE-SP, however, the relatively small detector size implies that a reconstructed muon belonging to an air shower containing multiple muons will appear as a ``single'' muon, since the rest of the muons from the same air shower will fall outside the detector acceptance. Given this, the interpretation of our measurement relies on the fact that ProtoDUNE-SP is located on the surface, and it samples mostly the low-energy muon flux. Simulation studies indicate that the mean muon energy after applying the selection cuts is approximately 6~GeV. We note that the results are similar when treating the period as a free parameter as well (the fitted amplitude is $5.0 \pm 1.5\%$ and period is $350 \pm 39 $), with the best fit value for the maximum muon rate being $46 \pm 29 $ days.
 
\ifmarkup
\par\smallskip\noindent{\color{red}\sout{[Table I -- compilation of seasonal-variation measurements from several experiments -- removed]}}\par\smallskip
\fi

\deleted{As we have previously discussed, the dominant cause of the seasonal variation effect for high-energy muons ($E_\mu > 10$ GeV) in underground detectors is the competition between the decay and interaction of secondary mesons ($\pi, K$) produced by the primary CR interactions. In the upper atmosphere, where most muons are produced, decay is slightly favored in the summer when the higher temperature causes the atmosphere to rise, leading to lower density and interaction rates. However, surface detectors, which primarily measure lower-energy muons ($E_\mu < 10$ GeV), observe a seasonal effect in the opposite direction due to muon decay. This is because low-energy muons are produced at higher altitudes under these conditions, increasing the probability that they decay in flight before reaching the detector. In the winter, a point of fixed pressure or overburden occurs at a lower altitude, so mesons are produced in a volume of higher density, favoring interactions.}
 
\added{The sign of the modulation depends on the muon energy because two effects compete. For high-energy muons ($E_\mu > 10$~GeV), as in deep underground detectors, the dominant effect is the competition between decay and interaction of the parent mesons. In summer the atmosphere is less dense at a given altitude, so more mesons decay and the rate is higher. In winter a point of fixed pressure lies at a lower altitude, so mesons are produced in denser air, which favors interaction and lowers the muon yield. Mesons decaying higher up also leave less energy for muon production lower down. Separately, for low-energy muons at the surface, the muon itself may decay in flight. In summer, muons of a few GeV are produced higher (typically 10--20~km), travel farther and decay more often before reaching the detector. This opposite effect dominates for our sample (mean energy $\approx 6$~GeV), giving a winter maximum. For details see Refs.~\cite{Gaisser2016,NOvA2026,Verpoest2024}.}
 
In order to simulate these effects, stemming from the fact that during the summer, the atmosphere expands, we consider the following approach\deleted{: since the total mass of the atmosphere does not change depending on the season, the expansion in the summer caused by the warmer temperature is vertical. In the approximation of an isothermal atmosphere, the ideal gas law implies that a $\pm 2\%$ change in the absolute temperature leads to a $\pm 2\%$ change in the altitude for a point of fixed pressure. Hence, we used the CORSIKA simulation with two different settings regarding the ``observation level'' variable (\cite{CORSIKAguide}, page 102, section 4.70) which defines the detector altitude, in order to approximate this expansion by an amount consistent with the approximate percentage change of atmospheric temperature.}\added{. In the approximation of an isothermal atmosphere with fixed surface pressure, the altitude of a point of fixed pressure is proportional to the absolute temperature, $h = H\ln(P_0/P)$ with $H = kT/(mg)$. A relative change $\delta$ of the absolute temperature therefore leads to the same relative change $\delta$ in that altitude.}
To quantify the atmospheric temperature modulation, we use data from the European Centre for Medium-Range Weather Forecasts (ECMWF)~\cite{dee2011}, which provides temperature measurements at multiple altitudes in the form of pressure levels. This dataset includes 37 discrete pressure levels, spanning from 1~hPa (upper atmosphere) to 1000~hPa (near the Earth's surface), allowing for a detailed reconstruction of the vertical temperature profile.
 
Defining an average atmospheric temperature as
\begin{equation}
T_{\mathrm{aver}} = \frac{\int_{P_1}^{P_2} T(P)\, dP}{\int_{P_1}^{P_2} dP},
\end{equation}
where $T(P)$ is the temperature at pressure level $P$, we evaluate the average temperature during periods corresponding to the maximum and minimum observed muon rates. We, thus, obtain $\sim$245~K in the winter and $\sim$ 256~K in the summer, i.e. a $4.3\%$ relative variation.

\begin{figure}[htbp]
    \captionsetup{
        format=plain,
        singlelinecheck=false,
        width=0.48\textwidth
    }
    \includegraphics[width=0.5\textwidth] {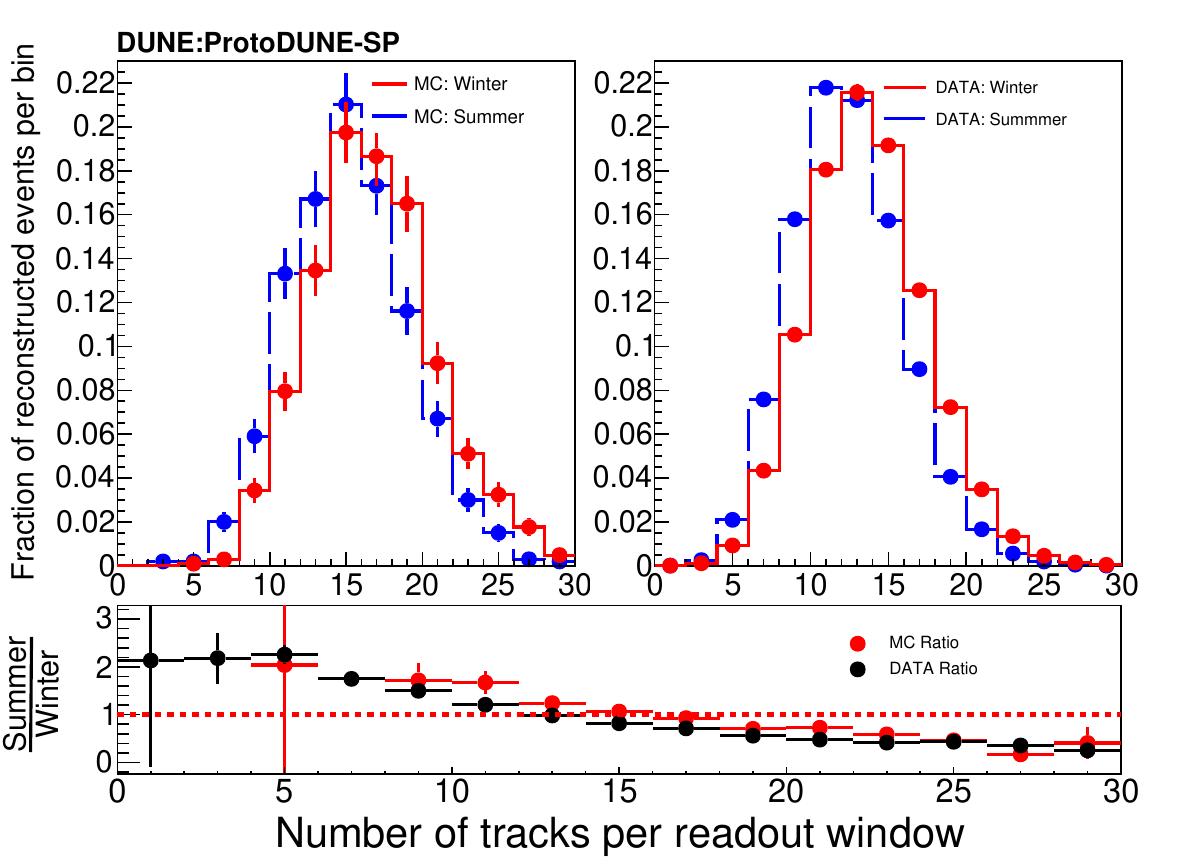}
    \caption{
            \justifying
    Number of muon tracks per event in summer (blue lines) and winter (red lines) for simulated (top left) and real (top right) events. Their ratio is also displayed (bottom) for real (black points) and simulated (red points) data, showing agreement.}
    \label{seasonall}
\end{figure}

\deleted{Then, using an averaged atmospheric density profile, we generate simulated samples for two different observation-level configurations differing by this $4.3\%$. These simulated events, analyzed in exactly the same way as the data, indicate a summer-to-winter muon rate variation of $(9.4 \pm 1.1)\%$, where the uncertainty is purely statistical. This result is in excellent agreement with the observed effect in our data, as also shown in Fig.~\ref{seasonall}.} \added{To illustrate the effect of the longer muon flight path in summer, we generated simulated samples with CORSIKA~\cite{CORSIKAguide} for two observation levels differing by $4.3\%$ of a typical muon production height of about 15~km, i.e.\ 455~m and 1100~m. Analysed in the same way as the data, they indicate a summer-to-winter variation of the muon rate of $(9.4\pm 1.1)\%$ (statistical uncertainty only), of the same order as the measured value of $(9.8\pm 2.8)\%$. This simplified treatment captures decay in flight, but not changes in the density profile and overburden or in meson interactions, and the result depends on the atmospheric model (two built-in CORSIKA profiles differ by about 4\% in the total rate). We therefore regard the comparison as qualitative.} Figure~\ref{seasonall} shows the distributions of the number of CR muon\added{ track}s per \deleted{event}\added{readout window} under summer and winter conditions, for simulated (left) and real data (right), along with their ratio (bottom). The ratio of summer to winter distributions for both data and simulation shows a \deleted{consistent}\added{similar} behavior\deleted{, demonstrating that the seasonal dependence of the muon multiplicity is accurately reproduced by the simulation}\added{; the simulated ratio is slightly higher than in data at low track numbers, which we attribute to the approximate seasonal treatment in the simulation}.
 
\begin{figure}[htbp]
    \captionsetup{
        format=plain,
        singlelinecheck=false,
        width=0.48\textwidth
    }
    \includegraphics[width=0.5\textwidth]{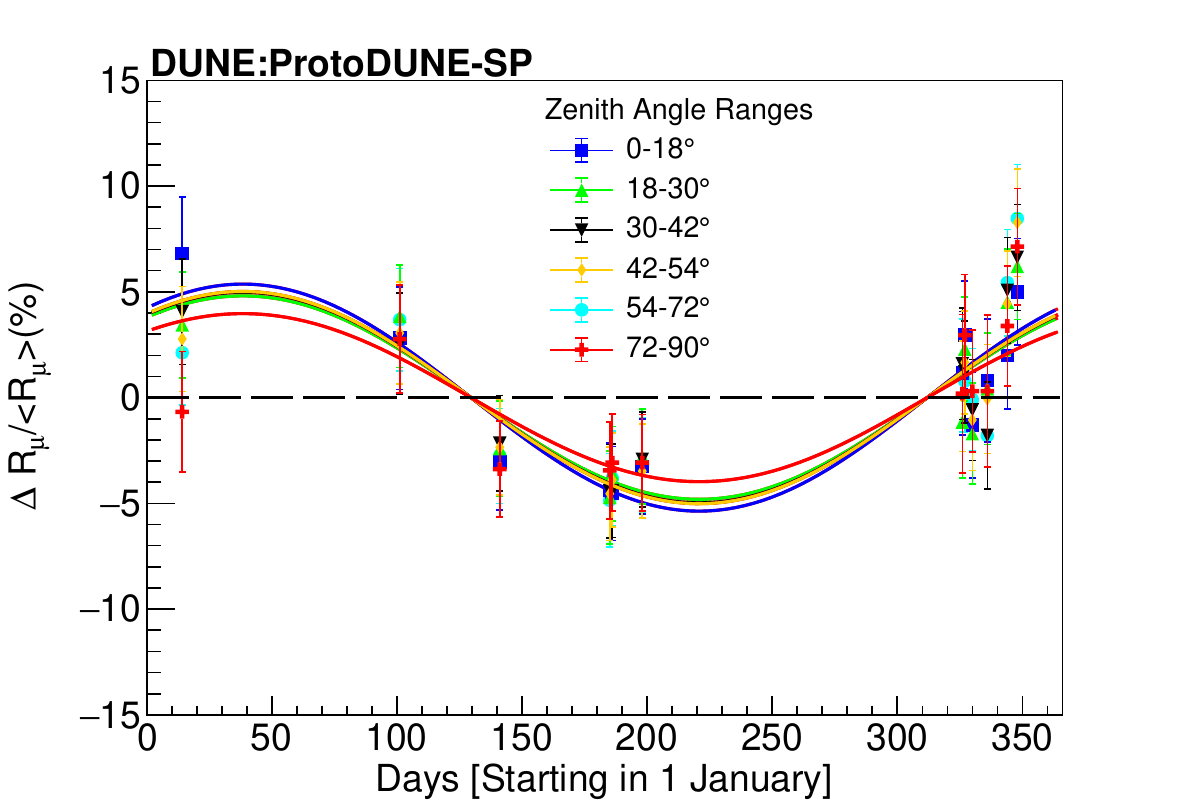}
    \caption{
            \justifying
    The percent change in the muon rate as a function of time (day of the year), fitted under the seasonal variation hypothesis for different zenith angle ranges with the phase held fixed: 0--18$^{\circ}$  (blue), 18--30$^{\circ}$  (green), 30--42$^{\circ}$  (black), 42--54$^{\circ}$  (yellow), 54--72$^{\circ}$  (cyan), and 72--90$^{\circ}$  (red). The error bars represent the quadratic sum of statistical and systematic uncertainties.}
    \label{seasonal_z}
\end{figure}

\deleted{Experiments such as NOvA \cite{NOvA:2021lsm} and MINOS \cite{Adamson_2014} have investigated the seasonal variation of the CR muon rate as a function of the muon zenith angle. We perform an analogous study, as shown in Fig.} \added{To check whether the modulation depends on the muon direction, we repeat the fit in six zenith-angle ranges with the phase held fixed, as shown in Fig.}~\ref{seasonal_z}, which presents the muon rate as a function of time (with fixed phase), together with the corresponding sinusoidal fits for several zenith angle ranges. The binning in zenith angle is chosen to ensure comparable statistical power across intervals. The amplitudes extracted from the sinusoidal fits for the different zenith angle ranges are reported in Table~\ref{tab:seasonal_variation1}. The results are consistent \deleted{across all zenith angle intervals within the quoted uncertainties; however, a consistent trend of a decrease in the amplitude vs zenith angle is seen, with an overall $\sim$1 $\sigma$ decrease observed when comparing small $(0-18^{\circ})$  with large $(72-90^{\circ})$ zenith angles. This is qualitatively consistent with results from the NOvA experiment \cite{NOvA:2021lsm}, which observes a decrease in average muon multiplicity for showers at large zenith angles. This trend is generally attributed to the higher typical muon energies and increased atmospheric path lengths rather than detector acceptance effects.}\added{with a constant amplitude across all zenith angle intervals within the quoted uncertainties.}

\begin{table}[htbp]
\footnotesize
\renewcommand{\arraystretch}{1.2}
\begin{tabular}{|c|c|}
\hline
\textbf{Zenith Angle Range} & \textbf{Amplitude (\%)} \\ \hline
\hline
$0-18^{\circ}$   & $5.4 \pm 1.2$  \\ \hline
$18-30^{\circ}$  & $4.8 \pm 1.1$  \\ \hline
$30-42^{\circ}$  & $5.0 \pm 1.1$  \\ \hline
$42-54^{\circ}$  & $5.0 \pm 1.1$  \\ \hline
$54-72^{\circ}$  & $5.0 \pm 1.1$  \\ \hline
$72-90^{\circ}$  & $4.0 \pm 1.2$  \\ \hline
\hline
\end{tabular}
\captionsetup{
        format=plain,
        singlelinecheck=false,
        width=0.48\textwidth
    }
\caption{
 \justifying
Amplitude as a function of the muon zenith angle with the phase fixed. The ranges are ordered from vertical to horizontal.}
\label{tab:seasonal_variation1}
\end{table}
\section{Conclusions}

In this paper, we report on the measurement of the cosmic muon rate seasonal variation with the ProtoDUNE-SP detector at the CERN Neutrino Platform with data collected from December 2018 to July 2020. After a set of dedicated run, event, and track selection criteria are applied, we observe a seasonal variation in CR muon rate at the 4 $\sigma$ level\added{, confirmed with pseudo-experiments, with an amplitude of $(4.9\pm 1.4)\%$ and a maximum in winter}. This result is consistent with \deleted{those obtained from previous experiments, and with models attributing such variations to atmospheric expansion and meson decay dynamics}\added{expectations for surface detectors sensitive to low-energy muons, where decay in flight dominates}. \deleted{Simulations incorporating temperature-driven atmospheric profiles reproduce the observed modulation, further validating this interpretation.} \added{A simplified simulation of the seasonal change of the muon flight path gives a modulation consistent in sign and approximate size with the measurement.} Beyond confirming \deleted{theoretical expectations}\added{qualitative expectations from atmospheric physics and simulation}, this result illustrates the broader scientific potential of surface liquid argon detectors, not only as prototypes but as sensitive instruments capable of probing subtle effects, such as percent-level seasonal variations in the muon flux. This builds confidence that the DUNE far detectors can operate stably over the multiple years that will be required to measure effects associated with neutrino CP violation and other phenomena that form the DUNE physics program.
 
\section{Acknowledgements}
 
The ProtoDUNE-SP detector was constructed and operated on the CERN Neutrino Platform. We
thank the CERN management for providing the infrastructure for this experiment and gratefully
acknowledge the support of the CERN EP, BE, TE, EN, IT and IPT Departments for NP04/ProtoDUNE-SP. This document was prepared by the DUNE collaboration using the resources of the Fermi
National Accelerator Laboratory (Fermilab), a U.S. Department of Energy, Office of Science, HEP
User Facility. Fermilab is managed by Fermi Research Alliance, LLC (FRA), acting under Contract
No. DE-AC02-07CH11359. This work was supported by CNPq, FAPERJ, FAPEG and FAPESP,
Brazil; CFI, IPP and NSERC, Canada; CERN; MŠMT, Czech Republic; ERDF, H2020-EU and
MSCA, European Union; CNRS/IN2P3 and CEA, France; INFN, Italy; FCT, Portugal; NRF, South
Korea; CAM, Fundación “La Caixa” and MICINN, Spain; SERI and SNSF, Switzerland; TÜBİTAK,
Turkey; The Royal Society and UKRI/STFC, United Kingdom; DOE and NSF, United States of
America, the Hellenic Foundation for Research and Innovation (H.F.R.I.) under the “3rd Call for H.F.R.I. Research Projects to support Post-Doctoral Researchers” (Project Number: 7036). This research used resources of the National Energy Research Scientific Computing
Center (NERSC), a U.S. Department of Energy Office of Science User Facility operated under
Contract No. DE-AC02-05CH11231.
 
\clearpage
\newpage
 
\bibliographystyle{apsrev4-2}
\bibliography{reference}
 
\end{document}